\documentclass[twocolumn]{aa}
\usepackage{graphicx}
\usepackage{txfonts}
\usepackage{natbib}
\bibpunct{(}{)}{;}{a}{}{,}

\begin{document}

\title{ The HARPS search for southern extra-solar planets\thanks{Based
    on observations made with the HARPS instrument on the ESO 3.6 m
    telescope at La Silla Observatory under the GTO programme ID
    072.C-0488.}  } 
\subtitle{XI. Super-Earths (5\,\&\,8\,M$_{\oplus}$) in a 3-planet
  system}

\author{
S.~Udry\inst{1}
\and X.~Bonfils\inst{2}
\and X.~Delfosse\inst{3}
\and T.~Forveille\inst{3}
\and M.~Mayor\inst{1}
\and C.~Perrier\inst{3}
\and F.~Bouchy\inst{4}
\and C.~Lovis\inst{1}
\and F.~Pepe\inst{1}
\and D.~Queloz\inst{1}
\and J.-L.~Bertaux\inst{5}
}

\offprints{S. Udry}

\institute{ 
  Observatoire de Gen\`eve, Universit\'e de Gen\`eve, 51 ch.
  des Maillettes, 1290 Sauverny,  Switzerland\\
  \email{stephane.udry@obs.unige.ch} 
  \and Centro de Astronomia e Astrof\'{\i}sica da Universidade de
  Lisboa, Tapada da Ajuda, 1349-018 Lisboa, Portugal 
  \and Laboratoire d'Astrophysique, Observatoire de Grenoble,
  Universit\'e J.~Fourier, BP 53,
  F-38041 Grenoble, Cedex 9, France 
  \and Institut d'Astrophysique de Paris, CNRS, Universit\'e Pierre et
  Marie Curie, 98bis Bd Arago, 75014 Paris, France 
  \and Service d'A\'eronomie du CNRS/IPSL, Universit\'e de Versailles
  Saint-Quentin, BP3, 91371 Verri\`eres-le-Buisson, France 
}

\date{Received ; accepted To be inserted later}

\abstract{This Letter reports on the detection of two super-Earth
  planets in the Gl\,581 system, already known to harbour a hot
  Neptune. One of the planets has a mass of 5\,M$_{\oplus}$ and
  resides at the ``warm'' edge of the habitable zone of the star. It
  is thus the known exoplanet which most resembles our own Earth. The
  other planet has a 7.7\,M$_{\oplus}$ mass and orbits at 0.25\,AU
  from the star, close to the ``cold'' edge of the habitable zone.
  These two new light planets around an M3 dwarf further confirm the
  formerly tentative statistical trend for i) many more very low-mass
  planets being found around M dwarfs than around solar-type stars and
  ii) low-mass planets outnumbering Jovian planets around M dwarfs.

\keywords{ stars: individual: Gl\,581, stars: planetary systems --
techniques: radial velocities -- techniques: spectroscopy } }

\maketitle

\section{Introduction}
\label{Sect_Intro}

M dwarfs are of primary interest for planet-search programmes. First
of all, they extend the stellar parameters domain probed for planets.
For high precision radial-velocity planet searches, M dwarfs are
excellent targets as well, because the lower primary mass makes the
detection of very light planets easier than around solar-type stars.
In particular, Earth-mass planets around M dwarfs are within reach of
current high-precision radial-velocity planet-search programmes.
Furthermore, the habitable zones of M dwarfs reside much closer to
these stars (around 0.1\,AU) than for Sun-like stars. Habitable
terrestrial planets are thus detectable today. Such detections will
provide targets for future space missions looking for life tracers on
other planets, like the ESA Darwin and NASA TPF-C/I projects.  To find
such very light planets in the habitable zone of M dwarfs, our
consortium \citep{Mayor-2003:a} dedicates $\sim$10\% of the Guaranteed
Time Observations on HARPS to the precise radial-velocity monitoring
of some 100 nearby M dwarfs.

In this Letter we present the detection of two additional planets
orbiting Gl\,581, where we previously found a 1$^{st}$ close-in
Neptune-mass planet. The minimum mass of the 2$^{nd}$ new planet is
5.03~terrestrial mass, the lowest for any exoplanet to date, close to
the 5.5\,M$_{\oplus}$ of the microlensing candidate
OGLE-2005-BLG-390Lb \citep{Beaulieu-2006} found at a large separation
from another M dwarf. It resides at the inner edge of the habitable
zone of Gl\,581. The 3$^{rd}$ planet, at 0.25\,AU from the star, is
also in the super-Earth category (7.7\,M$_{\oplus}$), and is situated
close to the outer edge of the habitable zone of the system.
Section~2 briefly recalls some relevant properties of the parent star.
Section~3 describes the precise HARPS velocities and characterizes the
new planets. We also examine the possibility that the long-period
low-mass planet is actually an artefact of dark spots modulated by
rotation of the star, and conclude that this is unlikely.  The Letter
ends with conclusions.

\section{Stellar characteristics of  Gl\,581}
\label{Sect_Star}

The paper reporting the first Neptune-mass planet on a 5.36-d orbit
around Gl\,581 \citep{Bonfils-2005} describes the properties of the
star.
We will here just highlight those characteristics which are most
relevant for this paper:

i) Gl\,581 is one of the least active stars of our HARPS M-dwarf
sample. \citet{Bonfils-2005} checked that line shapes are stable down
to measurement precision through bisector measurements on the
cross-correlation functions, and the level of all chromospheric
activity indices similarly points toward a low activity.  Such indices
represent useful diagnostics of the stellar radial-velocity jitter
from rotational modulation of star spots or other active regions on
the stellar surface, though no quantitative relation has been
established for M dwarfs yet.  \citet{Bonfils-2007} used variations of
these indices to unveil a 35~days rotation period for Gl\,674, later
confirmed by a photometric campaign, but those of Gl\,581 do not
measurably vary. The low rotational velocity which we measure for
Gl\,581 ($v\sin{i}$\,$<$\,1\,kms$^{-1}$) would require large spots to
produce noticeable radial-velocity variations through line
asymmetries. Figure~1 of \citet{Bonfils-2007}, displaying the Ca[II]
line for Gl\,674 and Gl\,581, clearly demonstrates that Gl\,581 is
significantly less active than Gl\,674. It very probably has
proportionately smaller spots, and a longer rotational period than the
35~days of Gl\,674. We finally note that Gl\,876, hosting the close-in
7\,M$_{\oplus}$ planet \citep{Rivera-2005}, is more active than
Gl\,581.  Gl\,581 is thus expected to have a very low intrinsic
radial-velocity noise.

\begin{figure}[t!]
\centering
\includegraphics[width=\hsize]{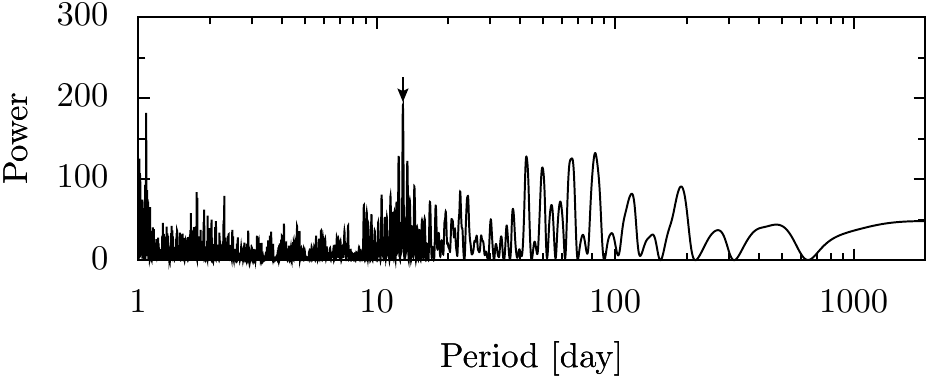}
\caption{ 
  Lomb-Scargle periodogram of the radial-velocity residuals around the
  1-planet solution, clearly showing a peak close to 13 days and some
  extra-power between 70 and 90 days. 
}
\label{udryf1}
\end{figure}

ii) Along the same lines, photometric observations of Gl\,581 depict a
stable star. \citet{Weis-1994} showed that on short time scales the
star varies by less than 6\,mmag, and the \citet{Lopez-2006}
photometric search for a potential transit of the 5.36-d period planet
similarly show a low 1.17\,mmag dispersion on scales of a few hours.
The Geneva photometry founds the star constant within the 5\,mmag
catalogue precision for 10.5 magnitude stars.  Photometric
observations have thus so far found no large spots. The photometric
stability will however need to be checked at high precision on longer
timescales.

iii) Very interestingly, Gl\,581 has a sub-solar metallicity
($[Fe/H]$\,=\,$-0.25$ in \citep{Bonfils-2005:b}; $[Fe/H]$\,=\,$-0.33$
in \citep{Bean-2006}), contrarily to most planet-host stars.
Mainstream theoretical and numerical studies of planet formation,
based on core-accretion models, predict that the joint effects of a
low-mass primary and low metallicity make giant planet formation very
unprobable. This finding is supported by the radial-velocity
observations \citep[e.g.][]{Santos-2004:b,Bonfils-2006}.  Formation of
low-mass planets on the other hand is not hampered for deficient
\citep{Ida-2004:b,Benz-2006} or low-mass stars
\citep{Laughlin-2004,Ida-2005}.

\section{Description of the Gl\,581 planetary system}
\subsection{HARPS radial-velocity measurements of Gl\,581}
\label{SectRV}

The 20 high resolution HARPS spectra available when we detected
Gl\,581\,b \citep{Bonfils-2005} have typical S/N per pixel of
$\sim$40, and at that time the typical radial-velocity uncertainty was
1.3\,ms$^{-1}$ per measurement, taking into account calibration
uncertainties.

\begin{figure}[t!]
\centering
\includegraphics[width=\hsize]{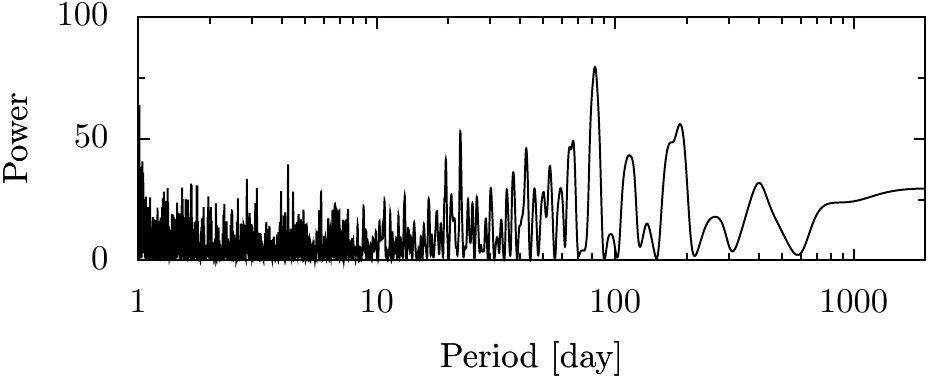}
\caption{Periodogram of the radial-velocity residuals around the 
  2-planet Keplerian model for Gl\,581 showing power at
  $P$\,=\,84\,d.}
\label{udryf2}
\end{figure}

The periodogram of their residuals from the 1-planet keplerian
solution showed a tentative 2$^{nd}$ signal at a frequency of
1/13\,d$^{-1}$. With this limited number of observations the low
amplitude of the 13-day velocity variation only had modest
significance, but it prompted us to gather 30 additional
high-precision observations with HARPS (uncertainty
$<1.5$\,ms$^{-1}$).  We also took advantage of a concerted effort to
improve the reduction pipeline with a special emphasis on wavelength
calibration \citep{Lovis-2007:a}.  These improvements are directly
visible on the new set of barycentric radial velocities (available in
electronic form at CDS): their average uncertainty is 0.9\,ms$^{-1}$
(including photon noise, calibration uncertainty and spectrograph
drift uncertainty).  The 50 high-precision HARPS radial velocities
confirm the 5.36-d period planet, and we now clearly see the 13~days
signal in the periodogram of the residuals around the 1-planet
solution (Fig.\,\ref{udryf1}).  Some power around 80 days is also
visible.  The false-alarm probability of the 13~days peak is only
0.0025.

\begin{figure}[t!]
\centering
\includegraphics[width=0.85\hsize]{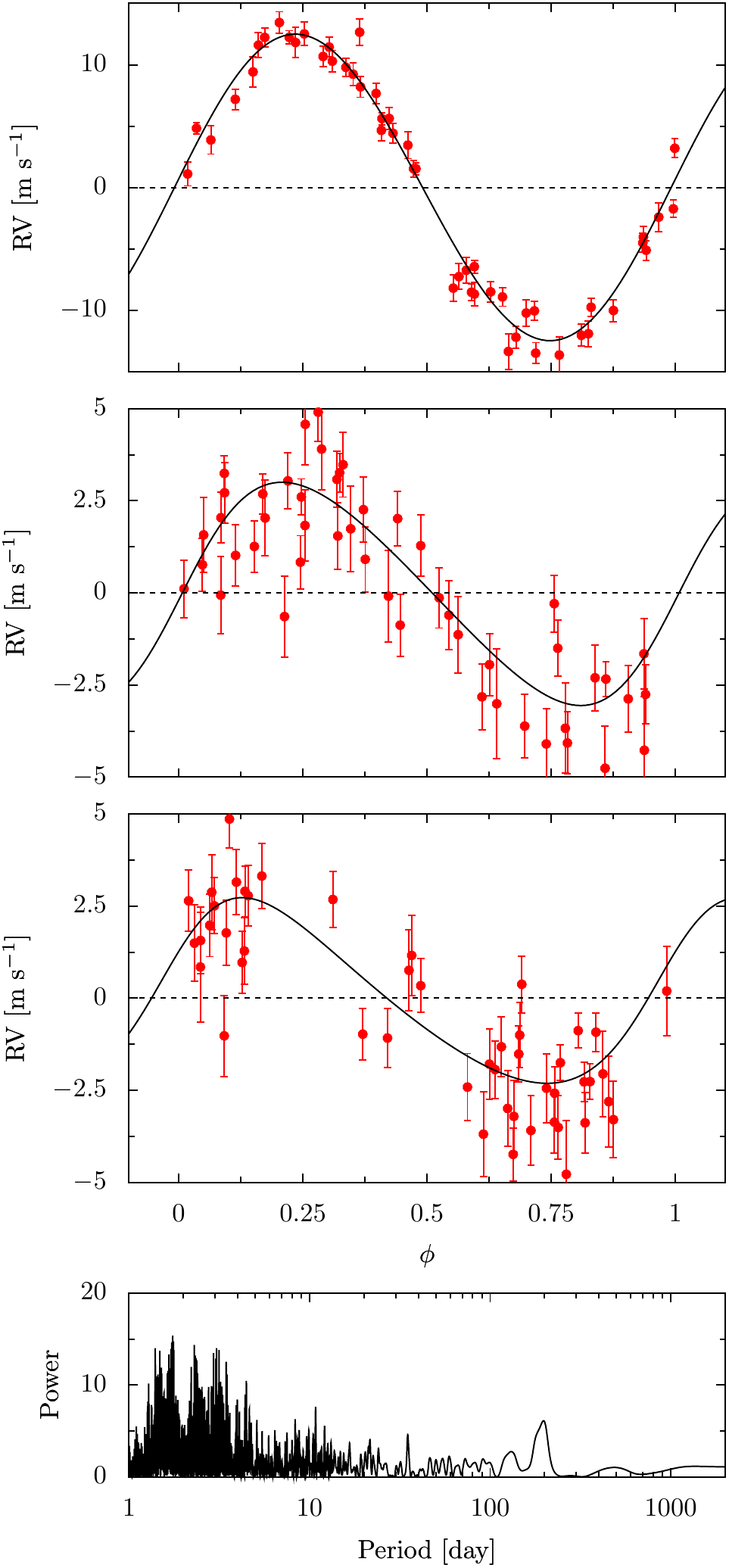}
\caption{3-planet Keplerian model of the Gl\,581 radial-velocity
  variations. The upper panels display the phase-folded curve of
  each of the planets, with points representing the observed radial
  velocities, after removing the effect of the other planets.    
  The bottom panel presents the periodogram of the residuals.}
\label{udryf3}
\end{figure}

\subsection{A 5\,M$_{\oplus}$ planet at 0.07\,AU from the star}

Although the new radial-velocity measurements strongly confirm the
5.36-day planet, their modeling by a single keplerian orbit is poor:
the residuals around the 1-planet solution are very high
(3.2\,ms$^{-1}$ standard deviation) compared with the 0.9\,ms$^{-1}$
typical measurement errors, and the reduced ${\chi}^2$ per degree of
freedom is $\chi^2_{red}$\,=\,17.3.  This, and the 13-days periodogram
peak, motivates investigating a 2-planet model. For the first planet,
that solution gives orbital parameters consistent with the
\citet{Bonfils-2005} orbit. The 2nd planet moves on a slightly
eccentric orbit ($e$\,$\simeq$\,0.28\,$\pm$\,0.06), with a period of
12.895~days.


The measured radial-velocity semi-amplitude is only 3.5\,ms$^{-1}$, or
4 times our typical noise on individual measurements. At this
small-amplitude radial-velocity variation, this solution represents a
highly significant improvement in the system modeling: $\chi^2_{red}$
decreases from 17.3 to 9.2, and the weighted rms of the residuals
around the solution is now 2.2\,ms$^{-1}$. We can note here that a
circular orbit for the 2nd planet provides a solution of equal quality
with a $\chi^2_{red}$\,=\,9.0. The observed dispersion of the
residuals is, however, still larger than the internal errors, and the
periodogram of the residuals from this 2-planet fit
(Fig.\,\ref{udryf2}) shows clear power at 84 days (the false alarm
probability of this signal is only 0.0028). In the next section we
examine this 3$^{rd}$ signal in terms of an additional planet, and
discuss whether it could instead be caused by magnetic activity.

\begin{figure}[t!]
\centering
\includegraphics[width=0.9\hsize]{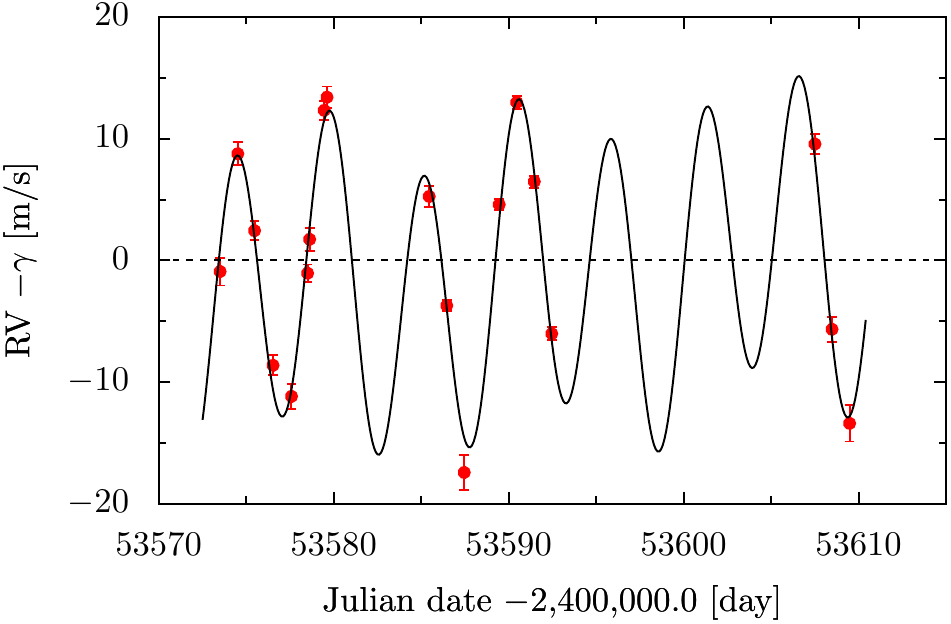}\\
\includegraphics[width=0.9\hsize]{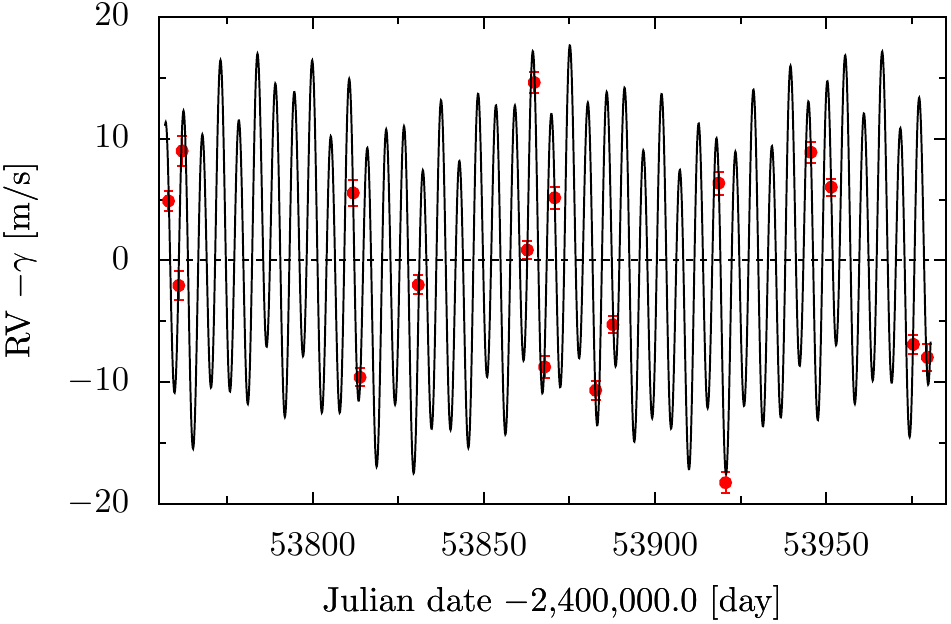}
\caption{Temporal display of the 3-planet Keplerian model of 
  Gl\,581, on time intervals with dense observational sampling.}
\label{udryf4}
\end{figure}

For the 0.31\,M$_{\odot}$ mass of Gl\,581 \citep{Bonfils-2005}, the
derived orbital parameters for the 2$^{nd}$ planet lead to a
$m_2\sin{i}$\,$\simeq$\,5.6\,M$_{\oplus}$ minimum mass and a
separation $a$\,=\,0.073\,AU. 
From the 0.013\,L$_{\odot}$ stellar luminosity
\citep{Bonfils-2005}, we compute an equilibrium temperature for the
planet of $-3^{\circ}$\,C (for a Venus-like albedo of 0.64) to
$+40^{\circ}$\,C (for an Earth-like albedo of 0.35). With a planetary
radius of $\sim$1.5\,R$_{\oplus}$ \citep{Valencia-2006} and a
temperature that would be $+20^{\circ}$\,C for a 0.5 albedo, Gl\,581c
is probably the most Earth-like of all known exoplanets. It is however
obvious that the actual surface temperature of the planet very much
depends on the highly uncertain composition and thickness of its
atmosphere, which govern both the planetary albedo and the strength of
the greenhouse effect. It is probable that the planet is located
towards the ``warm'' edge of the habitable zone around the star.  A
detailed study will also need to consider the possible tidal locking
of the planetary rotation to the orbital period.

\begin{table*}[t!]
\caption{Orbital and physical parameters derived from 
3-planet Keplerian models of Gl\,581 for the free-eccentricity
and circular cases. Uncertainties are directly derived from 
the covariance matrix.}
\label{udrytab3}
\centering
\begin{tabular}{ll|lccc|lccc}
\hline\hline
\multicolumn{2}{l|}{} & \multicolumn{4}{c|}{Circular case} 
                      & \multicolumn{4}{c}{Free eccentricity case}\\
\multicolumn{2}{l|}{\bf Parameter} &\hspace*{0mm} 
                     & \bf Gl\,581\,b & \bf Gl\,581\,c & \bf Gl\,581\,d
                     &\multicolumn{1}{c}{\hspace*{0mm}} 
                     & \bf Gl\,581\,b & \bf Gl\,581\,c & \bf Gl\,581\,d\\
\hline
$P$ & [days]      & & 5.3687$\pm$0.0003  &12.931$\pm$0.007  & 83.4$\pm$0.4  
                  & & 5.3683$\pm$0.0003  &12.932$\pm$0.007  & 83.6$\pm$0.7 \\
$T$ & [JD-2400000]& &52999.99$\pm$0.05 &52996.74$\pm$0.45 & 52954.1$\pm$3.7 
                  & &52998.76$\pm$0.62 &52993.38$\pm$0.96 & 52936.9$\pm$9.2 \\
$e$ &              & & 0.0  (fixed)  & 0.0 (fixed)   & 0.0 (fixed) 
                   & & 0.02$\pm$0.01 & 0.16$\pm$0.07 & 0.20$\pm$0.10\\
$V$ &[km s$^{-1}$] & & \multicolumn{3}{c|}{-9.2115 $\pm$ 0.0001}  
                   & & \multicolumn{3}{c}{-9.2116 $\pm$ 0.0002} \\
$\omega$ & [deg]   & & 0.0 (fixed)   & 0.0 (fixed)  & 0.0 (fixed) 
                   & & 273$\pm$42    & 267$\pm$24   & 295$\pm$28 \\
$K$ & [m s$^{-1}$] & & 12.42 $\pm$ 0.19   & 3.01$\pm$0.16  & 2.67$\pm$0.16
                   & & 12.48 $\pm$ 0.21   & 3.03$\pm$0.17  & 2.52$\pm$0.17\\
$a_1 \sin{i}$ & [10$^{-6}$ AU] & & 6.129  & 3.575  & 20.47
                               & & 6.156  & 3.557  & 18.98 \\
$f(m)$ &[10$^{-13} M_{\odot}$] & & 10.66  & 0.365  & 1.644
                               & & 10.80  & 0.359  & 1.305\\
$m_2 \sin{i}$ & [$M_{\mathrm{Jup}}$] & & 0.0490  & 0.0159  & 0.0263
                                     & & 0.0492  & 0.0158  & 0.0243 \\
$m_2 \sin{i}$ & [$M_{\oplus}$]       & & 15.6    & 5.06    & 8.3
                                     & & 15.7    & 5.03    & 7.7  \\
$a$           & [AU]                 & & 0.041   & 0.073   & 0.25
                                     & & 0.041   & 0.073   & 0.25\\  
\hline
$N_{\mathrm{meas}}$ &          & & \multicolumn{3}{c|}{50}
                               & & \multicolumn{3}{c}{50} \\
{\it Span}          & [days]   & & \multicolumn{3}{c|}{1050} 
                               & & \multicolumn{3}{c}{1050} \\
$\sigma$ (O-C) & [ms$^{-1}$]   & & \multicolumn{3}{c|}{1.28} 
                               & & \multicolumn{3}{c}{1.23} \\
$\chi^2_{\rm red}$ &           & & \multicolumn{3}{c|}{3.17}
                               & & \multicolumn{3}{c}{3.45} \\
\hline
\end{tabular}
\end{table*}

\subsection{A 3rd low-mass planet in the system}

Since the periodogram has significant power around 84 days, we
examined a 3-planet model. That solution changes the orbital
parameters of the inner two planets only slightly from the 2-planet
solution (lower eccentricities). The mass of the 2$^{nd}$ planet is
now 5.03\,M$_{\oplus}$. Adjusting their eccentricities finds that they
are not constrained according to a \citet{Lucy-71} test. We thus
provide in Table\,\ref{udrytab3} the orbital parameters for both cases
(free and fixed-to-zero eccentricities).  The 3$^{rd}$ planet has an
83.6~days period and a slightly eccentric orbit ($e$\,=\,0.2). The
inferred planet mass is 7.7\,M$_{\oplus}$ (8.3\,M$_{\oplus}$ in the
circular case) and the mean star-planet separation is 0.25\,AU,
putting the planet close to the outer edge of the habitable zone.
Aside from being most prominent in the frequency analysis, the 84-d
period naturally comes out in global solution searches based on the
genetic-algorithm approach. This makes a misidentified alias unlikely.
An on-going stability study of the system (Beust et al. in prep) shows
that the system is stable over millions of years, even in the more
eccentric case ($e+\sigma_e$).  Figure~\ref{udryf3} displays the
3-planet Keplerian solution together with the phase-folded radial
velocities, and Fig.\,\ref{udryf4} plots time sequences for densely
sampled measurement intervals.

Introducing a 3$^{rd}$ planet adds 5 free parameters and will thus
always lower residuals, but here the quality of the solution improves
impressively, and statistically very significantly: its $\chi^2_{red}$
drops from 9.2 to 3.45, and the 1.2\,ms$^{-1}$ rms residual is now
closer to the 0.9\,ms$^{-1}$ typical internal error. A modeling of the
system including planet-planet gravitational interactions gives the
same results and shows that for those small masses the mentioned
interactions are negligible.

Can the 84-d radial-velocity variation have another source?  Among the
very low-mass planets around M dwarfs, the recent Gl\,674\,b detection
provides a particularly illustrative comparison: the radial-velocity
measurements of Gl\,674 show two superimposed small-amplitude
variations with 4.7 and 35 days, but monitoring of chromospheric
indices and photometric observations demonstrate that the 35\,days
variation reflects rotational modulation of stellar activity, leaving
only one 11\,M$_{\oplus}$ planet with a 4.7-d period
\citep{Bonfils-2007}. This recent example emphasizes that the
interpretation of small-amplitude radial-velocity variations of M
dwarfs needs care, since most are expected to be at least moderately
active, and illustrates the value of chromospheric diagnostics and
photometric followups for these stars.

Since a comparison with Gl\,674 \citep[Fig.\,1 of][]{Bonfils-2007}
shows that Gl\,581 is significantly less active, its rotational period
is most likely longer than $\sim$40~days, and it could potentially
coincide with the 84-d signal. One therefore needs a serious look at
the possibility that the 84-d signal reflects a spot on the stellar
surface. At such a low rotation rate, one would however need a huge
spot to affect the radial velocities at the several $ms^{-1}$ level.
Scaling from \citet{Saar-1997}, a spot responsible for the observed
variation needs to cover 2.6\,\% of the stellar surface\footnote{ The
  same estimate for Gl\,674 with $P_{\rm rot}$\,=\,35~days,
  $R_{\star}$\,=\,0.29\,R$_{\odot}$, and $K$\,=\,5\,ms$^{-1}$ gives a
  1.7\,\% spot, close to the observed 2.6\,\%.}.  Such a large spot
would only be expected in a fairly active star, which Gl\,581 is not.
Planned spectroscopic (radial velocities and activity index) and
photometric monitoring of the star will settle that issue for good,
but we are already confident that the 3$^{rd}$ planet is real.

\section{Summary and discussion}

We report the detection of two new very light planets orbiting the low
metallicity M dwarf Gl\,581, already known to harbour a
15.7\,M$_{\oplus}$ closer-in planet \citep{Bonfils-2005}.  The high
radial-velocity precision reached with the HARPS spectrograph on the
ESO 3.6-m telescope enabled these discoveries.

The first planet, Gl\,581\,c, is a 5.03\,M$_{\oplus}$ super-Earth at a
distance of 0.073\,AU from the star. Its mass is the smallest found so
far for an exoplanet. At its separation from an M3 dwarf, the planet
resides at the inner edge of the habitable zone of this low luminosity
star.  With a radius close to 1.5\,R$_{\oplus}$, the planet is the
closest Earth twin to date.  The HARPS radial velocities also reveal a
longer-period planetary companion of mass 7.7\,M$_{\oplus}$, on a
83.6-d period orbit, close to the outer edge of the star habitable
zone. Considering uncertainties on the determination of the edges of
the habitable zone, mainly due to the lack of realistic cloud models,
these two planets are promising targets for future observatories. The
spectral characterization of their atmosphere would provide a crucial
constraint on the actual limits of the habitable zone.


The two new very low-mass planets further support statistical trends
already outlined in the literature:

i) Small planets (Neptune mass and below) are more frequent than giant
planets around M dwarfs (6 very low-mass detections against 3 Jovian
planets). This result was significant at the 97\,\% level before the
detection of the two new Gl\,581 planets \citep{Bonfils-2007}, even
without accounting for the poorer detection efficiency for lower-mass
planets.

ii) The fraction of detected Neptune (and lower-mass) planets around M
dwarfs is much larger than the corresponding ratio for solar-type
stars \citep{Bonfils-2006}. The absolute numbers of detections are
similar, but the number of surveyed solar-type stars is an order of
magnitude larger.  This may be an observational bias due to the lower
mass of M-dwarf primaries, or truly reflects more frequent formation
of Neptune-mass planets around M dwarfs. The factual conclusion
remains that Neptune-mass planets are easier to find around M dwarfs.

Recent planet-formation simulations \citep{Laughlin-2004,Ida-2005}
suggest that planet formation around low-mass primaries tends to
produce lower mass planets, in the Uranus/Neptune domain. Formation of
lower-mass planets is also favoured for solar-mass stars with
metal-poor protostellar nebulae
\citep{Ida-2004:b,Benz-2006}\footnote{Note, however, that there is no
  general consensus. \citet{Kornet-2006} suggest that smaller-mass
  primaries have denser disks, which would favour giant planet
  formation. Gravitational instability might also form super-Earth
  planets around M dwarfs as well \citep{Boss-2006}.}.  Gl\,581 is a
0.3\,M$_{\odot}$ metal-poor star, and its detected very light planets
are thus just what was expected around this star.  Additional
detections of very-low mass planets will help understanding these 2
converging effects.

From both our observational programmes and planet formation
simulations, very low-mass planets seem more frequent than the
previously found giant worlds.  They will thus provide preferential
targets for space photometric transit-search missions like COROT and
Kepler, and for projects like Darwin or TPF-I/C looking for biotracers
in the atmospheres of habitable planets.

\begin{acknowledgements}
  The authors thank the different observers from the other HARPS GTO
  sub-programmes who have also measured Gl\,581.  We especially thank
  Franck Selsis and Lisa Kaltenegger for thoughtful discussions,
  during the refering process, on the location of the habitable zone
  around Gl581.  We would like to thank the Swiss National Science
  Foundation (FNRS) for its continuous support to this project. XB
  acknowledges support from the Funda\c{c}\~ao para a Ci\^encia e a
  Tecnologia (Portugal) in the form of a fellowship (references
  SFRH/BPD/21710/2005).
\end{acknowledgements}

\bibliographystyle{aa}
\bibliography{udry_articles}

\newpage

\end{document}